# Effects of Oxygen Adsorption on the Surface State of Epitaxial Silicene on Ag(111)

Xun Xu[1], Jincheng Zhuang[1], Yi Du[1*], Haifeng Feng[1], Nian Zhang[2], Chen Liu[2], Tao Lei[2], Jiaou Wang[2*], Michelle Spencer[3], Tetsuya Morishita[4], Xiaolin Wang[1] & Shi Xue Dou[1]

[1] Institute for Superconducting and Electronic Materials (ISEM), University of Wollongong, Wollongong, NSW 2525, Australia

[2] Beijing Synchrotron Radiation Facility, Institute of High Energy Physics, Chinese Academy of Sciences, Beijing 100049, P. R. China

[3] School of Applied Sciences, RMIT University, Melbourne, VIC 3001, Australia

[4] Nanosystem Research Institute (NRI), National Institute of Advanced Industrial Science and Technology (AIST), Central 2,1-1-1 Umezono, Tsukuba, Ibaraki 305- 8568, Japan

* Correspondence and requests for materials should be addressed to Y. D. or J. W. (yi_du@uow.edu.au; wangjo@ihep.ac.cn)

## Abstract

Epitaxial silicene, which is one single layer of silicon atoms packed in a honeycomb structure, demonstrates a strong interaction with the substrate that dramatically affects its electronic structure. The role of electronic coupling in the chemical reactivity between the silicene and the substrate is still unclear so far, which is of great importance for functionalization of silicene layers. Here, we report the reconstructions and hybridized electronic structures of epitaxial 4×4 silicene on Ag(111), which are revealed by scanning tunneling microscopy and angle-resolved photoemission spectroscopy. The hybridization between Si and Ag results in a metallic surface state, which can gradually decay due to oxygen adsorption. X-ray photoemission spectroscopy confirms the decoupling of Si-Ag bonds after oxygen treatment as well as the relatively oxygen resistance of Ag(111) surface, in contrast to 4×4 silicene [with respect to Ag(111)]. First-principles calculations have confirmed the evolution of the electronic structure of silicene during oxidation. It has been verified experimentally and theoretically that the high chemical activity of 4×4 silicene is attributable to the Si $p_z$ state, while the Ag(111) substrate exhibits relatively inert chemical behavior.

Silicene, one single layer of silicon atoms packed in a honeycomb structure, has been predicted to be a new two-dimensional (2D) Dirac-fermion material[1-3]. The electrons in silicene behave as massless charge carriers that can exhibit transport at ultra-fast velocity, due to the linear energy-momentum dispersion relation at the Dirac point[4-6]. Strong spin-orbital coupling makes silicene a promising candidate material for the quantum spin Hall effect (QSHE)[5], and it is compatible with current Si-based device technologies. To date, epitaxial growth is the only method that can produce silicene on certain metal substrates[7-9]. The electronic structure of such epitaxial silicene is therefore significantly modified by a strong coupling with the substrate, which may annihilate silicene's Dirac-fermion characteristics[10]. On the other hand, epitaxial silicene in different superstructures may show various exotic physical and chemical properties due to new surface states that result from hybridization between Si and the metal substrate. For example, density functional theory (DFT) calculations have suggested that in 4×4 silicene [with respect to Ag(111)] on Ag(111), wave functions derived from the Si 3$p$ orbitals are delocalized into the substrate. The strong coupling, accompanied by the charge transfer from the substrate to the silicon, breaks the symmetry to modulate the band structure of 4×4 silicene. Consequently, a surface metallic band was observed in 4×4 silicene on Ag(111) by angle resolved photoemission spectroscopy (ARPES)[11]. This metallic band would lead to higher chemical reactivity of silicene in comparison to graphene, especially on the surface rather than the edges, which potentially facilitates functionalization of silicene. The chemical properties associated with such a hybrid electronic state in 4×4 silicene are still unclear, however, which is an obstacle to potential applications.

Oxygen adatoms can be used to probe and modulate local electronic states at the atomic level in 2D materials via the adsorption process, due to their high chemical activity. In graphene, it is well accepted that the local electronic structure can be tuned from the zero-gap state to a semiconducting state by changing the oxygen dose and the adsorption site of oxygen adatoms[12]. The interaction between oxygen adatoms and carbon atoms can reflect the intrinsic electronic properties of graphene. Similarly, silicene possesses high chemical reactivity to oxygen, which offers a feasible way to investigate its surface electronic structures by using the oxygen adatom as a chemical probe[1]. The hybrid surface metallic state in 4×4 silicene on Ag(111) would be perturbed by oxygen adatoms via the formation of covalent bonds between oxygen and silicon (or silver). Modulation of electronic states in silicene by oxygen adatoms is also expected, which is critical for the development of electronic devices.

In this work, we report the effects of oxygen adsorption on the surface state of epitaxial 4×4 silicene on Ag(111), as determined by scanning tunneling microscopy (STM) and ARPES at the atomic level. The hybridized surface metallic state is found to be highly sensitive to oxygen adatoms. It is revealed that the Shockley surface state on Ag(111) can be revived after the silicene is oxidized. Density functional theory (DFT) calculations indicate that the high chemical reactivity of 4×4 silicene originates from the Si $p_z$ state, which is in good agreement with the experimental results.

**Results**

Figure 1 shows 15 × 15 nm$^2$ STM topographical images of the Ag(111) surface and a 4×4 silicene monolayer epitaxially grown on the substrate. Quantum-interference patterns are clearly visible on the Ag(111) surface, as shown in Figure 1(a). Electrons in the two-dimensional surface states can be scattered by surface point defects, leading to periodic spatial oscillations of the electronic local density of states (LDOS)[13]. The LDOS can be used to identify a 2D electron gas, because interference will be observed if the 2D electron wave travels towards a scattering defect and encounters the backscattered wave[14]. In Figure 1(a), the quantum-interference pattern with a period of several tens of angstroms reflects the nature of the 2D electron wave in the Ag(111) surface-state band. In Figure 1(b), 4×4 silicene on a 1×1 Ag(111) surface exhibits a lattice constant of 1.06 nm. The low-buckled configuration can be verified by the different heights of the Si atoms at the edges. The height of buckling is 0.86 Å in 4×4 silicene, which is different from the calculated value for free-standing 1×1 silicene[15,16]. This is reasonable, because the metal passivation effect of the Ag(111) surface can affect the buckling of silicene through bonding between Si and Ag(111). We did not observe any quasiparticle interference pattern (QPI) in the as-grown silicene layer, which suggests the absence of Dirac fermion characteristics in epitaxial 4×4 silicene. In previous theoretical works, it was reported that the hybridization between Si and Ag leads to a symmetry breaking in 4×4 silicene, which, in turn, suppresses the Dirac quasiparticles[10,11]. The strong coupling, accompanied by the charge transfer, leads to a modulation of the electronic structure of silicene on Ag(111). In order to reveal the nature of the hybridization state, oxygen molecules were introduced onto the 4×4 silicene surface by a leak valve in precise doses at 77 K. Figure 1(c) and (d) shows typical STM images of silicene layers exposed to 10 Langmuir (L) and 600 L $O_2$, respectively. The topmost Si atoms in the buckled silicene are defined as "top-layer" (TL), and the other atomic layers are defined as "bottom-layer" (BL). At the low oxygen dose, the oxygen adatoms prefer to reside on a bridging site that forms a Si(TL)-O-Si(BL) configuration. The Si-O bonds significantly modulate the surface metallic band in silicene on Ag(111). As shown in the Figure 1(c) inset, a gapped electronic state was identified in scanning tunneling spectroscopy (STS) measurements carried out at oxygen adatom sites on the silicene layer. The surface metallic band is therefore tuned to a semiconducting-like characteristic. When the oxygen dose is increased up to 600 L, the silicene layer is oxidized and forms a disordered structure, as shown in Figure 1(d). Some areas of bare Ag(111) substrate were exposed. Interestingly, the QPI pattern again appears on the Ag(111) surface with the same oscillating period as observed on clean Ag(111) surfaces before the growth of silicene, indicating that Ag(111) substrate acts as an inert material compared to 4×4 silicene in the process of oxidization.

Figure 2 shows the ARPES results on occupied states along the $\Gamma$-$M_{Ag}$ and $\Gamma$-$K_{Ag}$ directions of 4×4 silicene/Ag(111) before and after oxidization. ARPES was conducted in order to reveal the details of the surface electronic structures, as well as the origin of the electronic hybridization between the silicene and the Ag(111) substrate. Figure 2(d) shows the reciprocal space Brillouin zones (BZ) of 1×1 Ag(111) (blue hexagon), free-standing silicene (dashed red hexagon), and 4×4 epitaxial silicene with respect to Ag(111) (or 3×3 silicene with respect to 1×1 silicene) (orange hexagons). Note that the $M_{Ag}$ and $K_{Ag}$ points of Ag(111) coincide with the $\Gamma_{Si}$ and $K_{Si}$ points of 4×4 silicene in the BZ. Figure 2(a) displays the Shockley surface state (SSS) of Ag(111) substrate at the BZ center $\Gamma$ point ($k = 0$ Å$^{-1}$). The SSS arises primarily from surface states of nearly free electrons and is associated with the special boundary conditions introduced by the metal/vacuum interface[17]. The typical bulk $sp$-band of Ag lies across the Fermi level[11] at $k = 1.15$ Å$^{-1}$. As the coverage of silicene increases, the SSS and $sp$-band of Ag become faint, and the SSS eventually disappears when the Ag(111) surface is fully covered by the silicene layer, as shown in Figure 2(b). The weak Ag $sp$-band is still observable, which indicates that this band remains stable upon Si deposition. There is a clear new "∩"-shaped state with a top point at the $M_{Ag}$ point. As shown in Figure 2(e), the "∩"-shaped state along the $\Gamma$-$K_{Ag}$ direction exhibits a variation, where the band traverses the Fermi surface at the $K_{Ag}$ ($K_{Si}$) point. The results of band structures along both the $\Gamma$-$M_{Ag}$ and the $\Gamma$-$K_{Ag}$ directions are consistent with previous reported works[11], indicating that the "∩"-shaped state is attributable to a hybridization of Si and Ag orbitals that resembles the $\pi$-band dispersion in graphene[11]. The apex of the state at $k = 1.28$ Å$^{-1}$ in Figure 2(b) is about 0.15 eV below the Fermi level, which is the saddle point of the surface state and at the middle point between two adjacent $K_{Ag}$ ($K_{Si}$) points. It should be noted that this feature is absent in the clean Ag(111) spectra and has been associated with a Ag(111)-related surface band that appears only after Si deposition. In recent studies, a different superstructure of silicene, namely, the √3×√3 reconstruction with respect to 1×1 silicene, was formed as a second layer on 4×4 silicene grown on Ag(111) substrate. The predicted Dirac cone has been observed in the √3×√3 silicene[18,19], which demonstrates that √3×√3 silicene is a true 2D Dirac-fermion material. Both ARPES and STS measurements indicated a strong electron doping (*n*-type) effect in multilayer √3×√3 silicene evoked by the Ag(111) substrate[6,18]. Consequently, the Fermi level in multilayer silicene is lifted by about 0.3 eV, even though there is no hybridized state that can be observed. Thus, the electronic structure of monolayer 4×4 silicene is completely different from that of multilayer √3×√3 silicene. High surface chemical activity would be expected in 4×4 silicene because the metallic HSB in 4×4 silicene leads to a lower work function compared with √3×√3 silicene. Figure 2(c) and (f) shows ARPES results on an oxidized 4×4 silicene/Ag(111) sample under an oxygen dose of 600 L along the $\Gamma$-$M_{Ag}$ and $\Gamma$-$K_{Ag}$ directions, respectively. The signature metallic HSB has disappeared. The two ARPES results show similar features to each other. In despite of the disappearance of the metallic HSB, the well-defined SSS bands are revived in the oxidized silicene/Ag(111) sample. The intensities of the Ag(111) states are weak because the Ag(111) surface is still partially covered by silicene oxide. Moreover, an asymmetric band with the highest energy level at about -0.6 eV can be observed in Figure 2(c). In order to elucidate the origin of this asymmetric band, we carried out STS measurements on areas of the Ag(111) surface and the silicene oxide in the same sample (not shown here). An energy gap of 1.2 eV was opened up on silicene oxide, while the Ag(111) showed typical metallic characteristics. Therefore, the asymmetric state is attributed to the valence band (VB) originating from silicene oxide. It is worth noting that SSS in metal is extremely surface sensitive, so that it can reflect modifications of surface atomic and electronic properties[20,21]. In our ARPES results, the revived SSS in the oxidized sample demonstrates that oxygen would preferentially react with Si rather than Ag(111) when a low oxygen dose is present. As a result, the surface states of Ag(111) can be chemically protected against oxygen molecules by 4×4 silicene. The disappearance of the HSB in Figure 2(c) and (f) implies that the hybridization between Si and Ag is broken due to oxidation.

In order to confirm the influence of oxygen adatoms on the hybridization between Ag(111) and 4×4 silicene, a detailed X-ray photoelectron spectroscopy (XPS) characterization of the chemical bonds in the samples was carried out. Figure 3(a) and (b) shows Ag 3$d$ core level XPS spectra for 4×4 silicene deposited on an Ag(111) sample before and after oxygen treatment, respectively. The experimental data points are displayed with black dots, while the fitted curves are red lines. For the bare Ag(111) substrate, the Ag 3$d_{3/2}$ and 3$d_{5/2}$ peaks at 371.5 eV and 365.5 eV have originated from Ag-Si bonds. A downward energy shift (~0.7 eV) for the Ag 3$d$ orbital is observed after the deposition of silicene, where the Ag-Si chemical bond forms, indicating that the chemical activity of silicene is higher than for the pure Ag-Ag bond arising from Ag$^0$. Peak splitting of the Ag 3$d$ line was observed after the exposure to 600 L oxygen in Figure 3(b). The peaks could be decomposed in two contributions, coming from the Ag-Ag bond component and the Ag-Si bond component, respectively. The dramatic fall in intensity of Ag-Si and the recovery of Ag-Ag bonds make it manifest that Ag-Si bonds are broken after the oxygen treatment. Moreover, no Ag-O chemical structure is present in the XPS spectrum, which implies that oxygen adatoms most likely only form bonds with silicon atoms, which supports the resurgence of QPI patterns on Ag(111) after oxidation, as shown in Figure 1(d). Figure 3(c) and (d) shows Si $2p$ core level spectra for the sample before and after oxygen treatment, respectively. The fitting results for the Si $2p$ line, as shown in Figure 3(c), make it clear that there are two groups of bonding components, labelled as Si1 and Si2, respectively. The energy gap of the two peaks in each group is a constant value, indicating that the two fitting peaks in one group are related to two Si $2p_{3/2}$ and $2p_{1/2}$ peaks, respectively. The Si2 peaks at a binding energy around 98.8 eV are related to the Si-Si in silicene, consistent with the previous report[22]. The Si1 group is attributed to Si-Ag bonding, since there are no other elements induced in the process of deposition, combined with the fitting results on Ag-Si bonding in Figure 3(a) and (b). The Si-O peaks clearly present after oxygen treatment. The binding energy value (101.6 eV) is lower than the peak position of $SiO_2$-like binding energy (102.3 eV)[23], indicating that the valence states of Si-O bonds are lower than $Si^{4+}$. Therefore, silicene was not fully oxidized to $SiO_2$, which is consistent with previous report[24] in which the oxygen adatoms are the most energetically favored on the surface of silicene. The intensity of the peaks related to the Si-Ag bonds is significantly reduced with the emergence of the Si-O peak. The variation of the peak intensity demonstrates that oxygen adatoms prefer to decouple the Si-Ag bonds rather than the Si-Si bonds. The XPS results are in a good agreement with our STM and ARPES results, and confirm the decoupling of Si-Ag bonds after oxygen treatment, as well as the relatively high oxygen resistance of the Ag(111) surface, in contrast to 4×4 silicene.

Finally, we carried out density functional theory (DFT) calculations to investigate the revived SSS on Ag(111) and to confirm the origins of the VB in silicene oxide, as shown in Figure 4. The first step in our calculation was to determine the superstructure of silicene grown on Ag(111). One layer of silicene was put on top of 5 layers of 4×4 Ag(111). The simulated structure shows the same reconstruction as in our STM results, as displayed in Figure 4(a). The Ag $d$-state and the Si $p$-state make the heaviest contributions to the density of states (DOS) at the Fermi level ($E_F$), which indicates that the metallic HSB is indeed contributed by the $p_z$ electrons of Si atoms and the 4$d$ electrons of the Ag(111) substrate. We then put 0.5 monolayer (ML) oxygen on the stabilized silicene surface. After running molecular dynamics for 7 ps, all the singly coordinated Si atoms have moved to bridge sites, indicating that there is an energy barrier that needs to be overcome for the other O atoms to move to more highly coordinated sites on silicene. Meanwhile, the Si atomic layer becomes disordered, demonstrating that the silicene oxide layer has started to decouple from the underlying Ag(111) substrate. The disordering of the Si overlayer induced by oxygen adatoms is in excellent accordance with the STM observations on the disordered nature of silicene oxide. Figure 4(c) shows the calculated DOS on 4×4 silicene covered by 0.5 ML oxygen. The deep level ($< -2$ eV) is mainly contributed by Ag $d$-states. The DOS near $E_F$, however, consists of Ag, Si, and O orbitals, as shown in the inset of Figure 4(c). It should be pointed out that the Si 3$p$ states and O 2$p$ states form a new band below $E_F$, although Si and O also contribute some partial DOS at $E_F$. The top of this band is at -0.4 eV, which matches well with the asymmetric band (-0.6 eV) shown in the ARPES results [Figure 2(c, f)]. Thus, we confirmed that this shallow band is the VB of partially oxidized silicene. It should be noted that 0.5 ML oxygen is not enough to oxidize the whole silicene layer, so that the hybridization between Si and Ag still exists in some regions. For areas of silicene oxide, the Shockley surface state would be revived due to decoupling between the silicene overlayer and the Ag(111). The Ag state at the Fermi level could be contributed by both the metallic HSB and the revived Shockley surface state.

**Discussion**

As mentioned above, 4×4 silicene on Ag(111) demonstrates high chemical reactivity towards oxygen, which is expected to be utilized to further functionalize epitaxial silicene layers. The binding energy between the epitaxial silicene layer and the Ag(111) surface is approximately 0.7 eV[25], which is much smaller than the binding energy for Si-O (between 4.0 and 12.0 eV[26]). The oxygen thus tends to bond firstly with Si atoms in the silicene instead of Ag atoms in the substrate. Moreover, the energy required for oxygen adsorption on Ag(111) is much higher than on the Si surface with dangling bonds ($p_z$ orbital), and therefore, bare Ag(111) surface but not silver oxide appears in the fully oxidized silicene sample[25,27]. Due to the characteristic $sp^3$ hybridization of Si, energetically stable Si-O-Si bonds would be expected when silicene is exposed to a high oxygen dose (600 L). The Si-O bond length in oxidized 4×4 silicene on Ag(111) is between 1.63 Å and 1.67 Å, as derived from our DFT calculations, which is somewhat smaller than the bond length of about 1.70 Å in $SiO_2$. Consequently, the silicene layer crumples during oxidation, resulting in some "silicene-free" areas. Our theoretical and experimental results also suggest that the silicene oxide layer could possibly detach from the Ag(111) substrate and form quasi-freestanding nanosheets. By analogy with graphene, it is proposed that reducing these quasi-freestanding nanosheets may offer a feasible way to obtain freestanding silicene nanosheets [or reduced silicene oxide (RSO)].

In conclusion, we have investigated the oxidation effects on the structure and electronic properties of 4×4 silicene on Ag(111). After oxidation, the silicene oxide exhibits a disordered structure with a semiconductor-like band structure. By combining DFT calculations and ARPES results, it is verified that the 2D metallic surface state in 4×4 silicene on Ag(111) is attributable to hybridization of Si $p_z$ and Ag $d$ states. The hybridization is broken and the Ag(111) Shockley surface state can be revived when the silicene is oxidized. This surface band demonstrates high chemical activity, which will facilitate chemical functionalization of silicene layers.

**Methods**

**Sample preparation.** The silicene layers were grown on Ag(111) substrate by the deposition of silicon from a heated silicon wafer in a preparation chamber attached to a low temperature (LT)-STM system under ultra-high vacuum (UHV, $< 5 \times 10^{-11}$ torr). A clean Ag(111) substrate was prepared by Ar$^+$ sputtering followed by annealing at 550$^\circ$C for several cycles. The deposition flux of Si was 0.08 monolayers per minute (ML/min). The temperature of the Ag(111) substrate was kept at 220$^\circ$C during deposition. Oxygen molecules were introduced onto the silicene surface by a leak valve. The Langmuir (L) is used as the unit of exposure of $O_2$, *i.e.* 1 L is an exposure of $10^{-6}$ torr $O_2$ in one second.

**STM and STS characterizations.** The STM and STS measurements were carried out on an ultra-high vacuum (UHV) LT-STM system (SNOM1400, Unisoku Co.) in UHV ($< 8 \times 10^{-11}$ torr) at 77 K. STS differential conductance ($dI/dV$) (where $I$ is current and $V$ is voltage) measurements were conducted with lock-in detection by applying a small modulation of 20 mV to the tunnel voltage at 973 Hz. Before STS measurements, the Pt/Ir tip was calibrated on a silver surface.

**ARPES and X-ray photoemission spectroscopy (XPS) characterizations.** *In-situ* ARPES and XPS characterizations were performed at a photoelectron spectroscopy station in the Beijing Synchrotron Radiation Facility (BSRF) using a SCIENTA R4000 analyzer. A monochromatized He-I light source (21.2 eV) was used for the band dispersion measurements. The total energy resolution was set to 15 meV, and the angular resolution was set to ~0.3°, which gives momentum resolution of ~0.01 Å$^{-1}$. The XPS experiments were performed at Beamline 4B9B, and the variable photon energies used were referenced to a fresh Au polycrystalline film. Photons at 700 eV, 500 eV, and 180 eV were used to excite the Ag-3$d$ and Si-2$p$ electrons in the samples, and the total energy resolutions were about 0.4 meV, 0.3 meV, and 0.15 meV, respectively. All the XPS data are fitted using the *XPS Peak* 4.1 software package. All the background subtraction was calculated by the "Shirley + Linear" background approach. All the XPS peaks were fitted by Gaussian-Lorentzian functions. The silicene samples used in the ARPES and XPS characterizations were prepared under the same conditions as for the STM and STS characterizations.

**DFT calculation details.** We performed density functional theory calculations and *ab initio* molecular dynamics (AIMD) simulations using the Vienna *Ab initio* Simulation Package (VASP)[28-30]. The exchange-correlation Perdew−Burke−Ernzerhof functional and the ion−electron interaction as described by the projector augmented wave method were used[31,32]. A plane-wave basis set with an energy cut-off of 400 eV was used with a Monkhorst-Pack k-point mesh of 13 × 13 × 1 for the geometry optimisation and the Γ point for the AIMD. The 4×4 silicene/Ag(111) system was modelled using the details published previously[33]. To model adsorption of oxygen in this system, we initially placed oxygen atoms in top sites above the Si atoms at ½ ML coverage. A geometry optimisation was performed using medium precision, followed by high precision. During the optimisation the bottom 2 layers of Ag atoms were kept fixed, while all other atoms were allowed to relax until the total energy converged to $< 10^{-4}$ eV, and the Hellmann–Feynman force on each of the atoms was allowed to relax to $< 0.03$ eVÅ$^{-1}$. An *ab-initio* molecular dynamics simulation of 7 ps, using a time step of 1 fs, followed by a subsequent partial geometry optimisation, showed that the Si layer becomes disordered and starts to delaminate from the Ag surface to form a silicon oxide type structure.

**Acknowledgements**

This work was supported by the Australian Research Council (ARC) through a Discovery Project (DP 140102581) and Linkage Infrastructure, Equipment and Facilities (LIEF) grants (LE100100081 and LE110100099). Y. Du would like to acknowledge support by the University of Wollongong through the Vice Chancellor's Postdoctoral Research Fellowship Scheme and a University Research Council (URC) Small Grant. The authors also acknowledge the valuable support and assistance from Dr T. Silver.

**Author Contributions**

X. X., J. Z. and Y. D., did the sample preparations and STM characterizations. N. Z., C. L., T. L., and J. W. did ARPES and XPS measurements. M. S. and T. M. performed DFT calculations. X. W., S. X. D. and H. F. helped with data analysis. X. X., J. Z. and Y. D. wrote the paper.

**Additional Information**

The authors declare no competing financial interests.

**Figures Captions**

**Figure 1.** Topographical images of Ag(111) substrate and 4×4 silicene grown on Ag(111): (a) STM topographical image of clean Ag(111) substrate, which shows a clear quantum-interference pattern due to point defects (scanning area 15 nm × 15 nm, $V_{bias}$ = -0.2 V, $I$ = 4 nA). (b) STM topographical image of 4×4 silicene on Ag(111) (scanning area 15 nm × 15 nm, $V_{bias}$ = -0.8 V, $I$ = 2 nA). Inset is an enlarged view of an area 4 nm × 4 nm in size. (c) STM image of silicene layer oxidized by an oxygen dose of 10 L. O adatoms prefer to reside at bridge sites. The inset contains STS spectra of silicene and silicene oxide samples, indicating that there is gap opening due to oxidation. (d) STM image of the 4×4 silicene oxidized under 600 L $O_2$. The bare Ag(111) surface can be seen at the bottom left of (d).

**Figure 2.** Energy vs. *k* dispersion measured by ARPES for (a) clean Ag(111) surface, (b) 4×4 silicene grown on Ag(111) along the $\Gamma$-$M_{Ag}$ direction, and (c) oxidized silicene on Ag(111) along the $\Gamma$-$M_{Ag}$ direction, respectively. SSS in (a) and (b) denotes the Shockley surface state. HSB in (b) denotes the hybrid surface band. (d) Schematic diagram of the BZ for 4×4 silicene grown on Ag(111): red, blue and orange honeycomb structures correspond to free-standing (FS) silicene, Ag(111) and 4×4 silicene with respect to Ag(111) (or 3×3 silicene with respect to 1×1 silicene), respectively. (e) 4×4 silicene grown on Ag(111) along the $\Gamma$-$K_{Ag}$ direction, and (f) oxidized silicene on Ag(111) along the $\Gamma$-$K_{Ag}$ direction, respectively.

**Figure 3.** Representative Ag 3*d* core level XPS spectra of 4×4 silicene on Ag(111) (a) before and (b) after oxidation. Si 2*p* core level XPS spectra of 4×4 silicene on Ag(111) (c) before and (d) after oxidation. Si1 and Si2 represent Si-Ag and Si-Si, respectively. The spectra indicate that the 4×4 silicene layer is oxidized and decoupled from Ag(111) when the oxygen dose is as high as 600 L.

**Figure 4.** (a) DFT-simulated structure of 4×4 silicene on Ag(111) substrate. (b) Initial adsorption sites of oxygen adatoms on silicene. (c) Calculated DOS of oxidized silicene with oxygen coverage of ½ ML. The inset is an enlarged view of the DOS near the Fermi level. (d) The energy-favoured stable adsorption site after running molecular dynamics for 7 ps. It indicates that oxygen adatoms prefer to form Si-O-Si bonds at bridge sites in the 4×4 silicene surface. Red, yellow, and blue balls in (b) and (d) represent oxygen, silicon, and silver atoms, respectively.

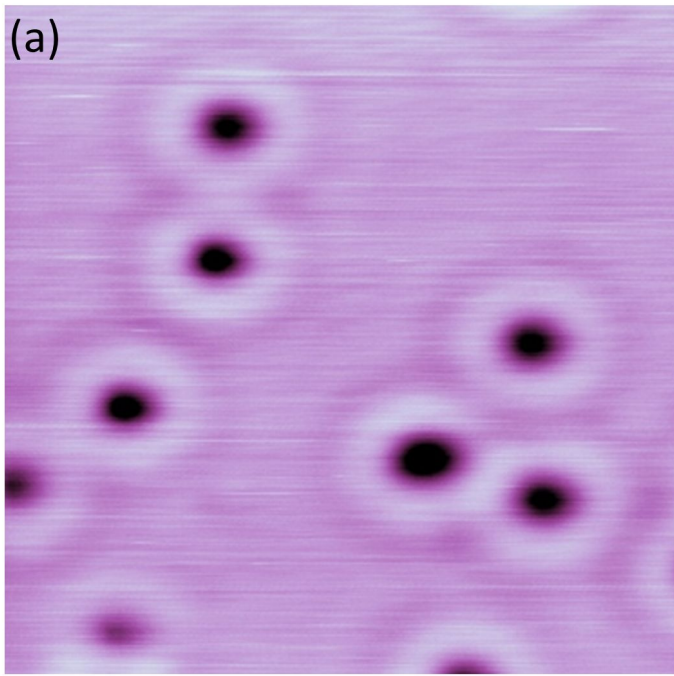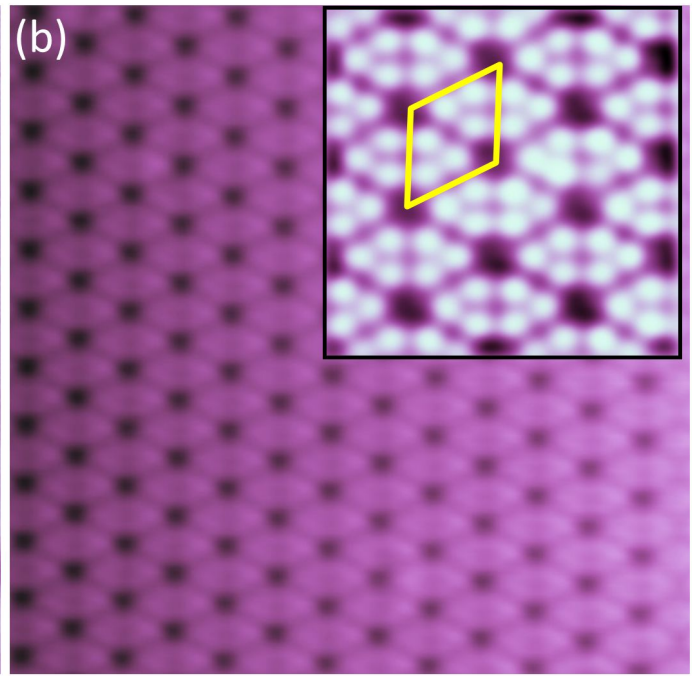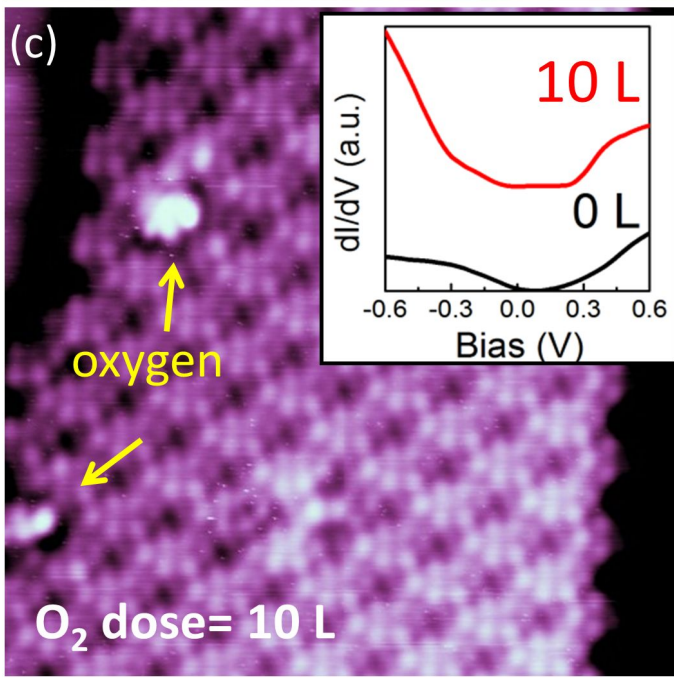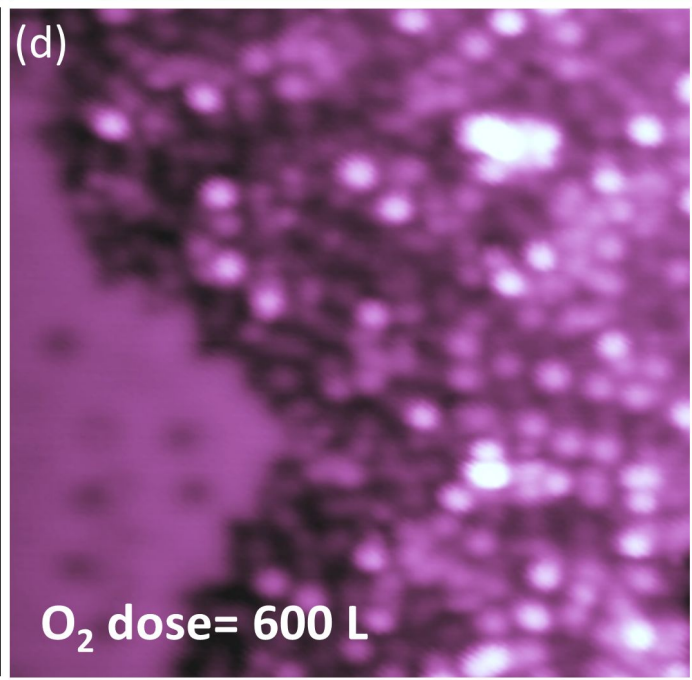

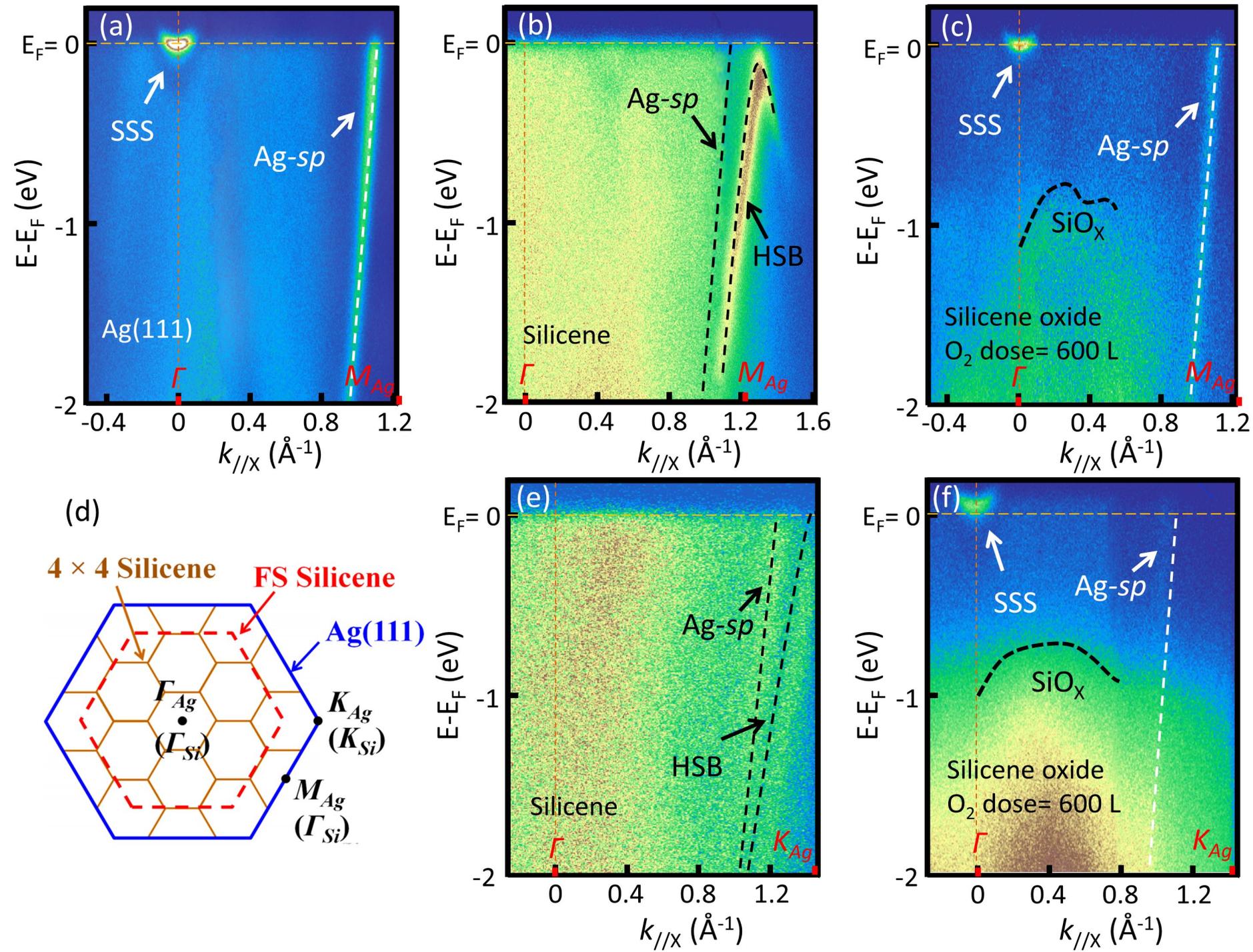

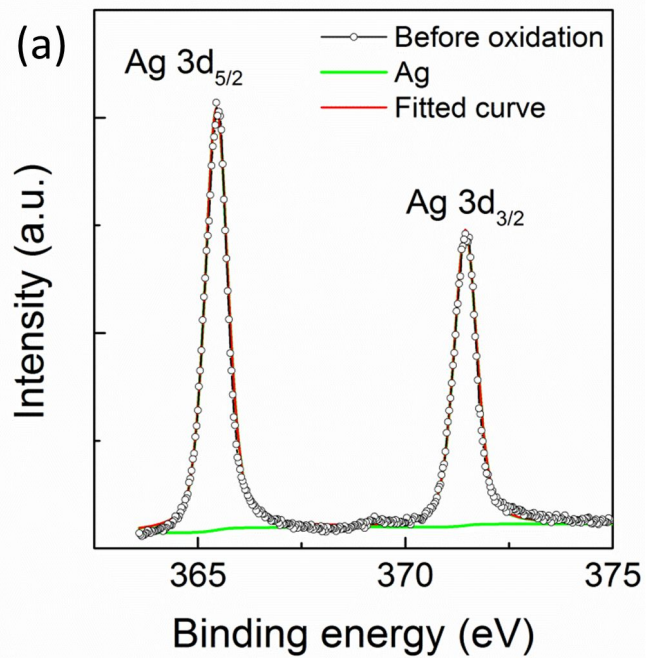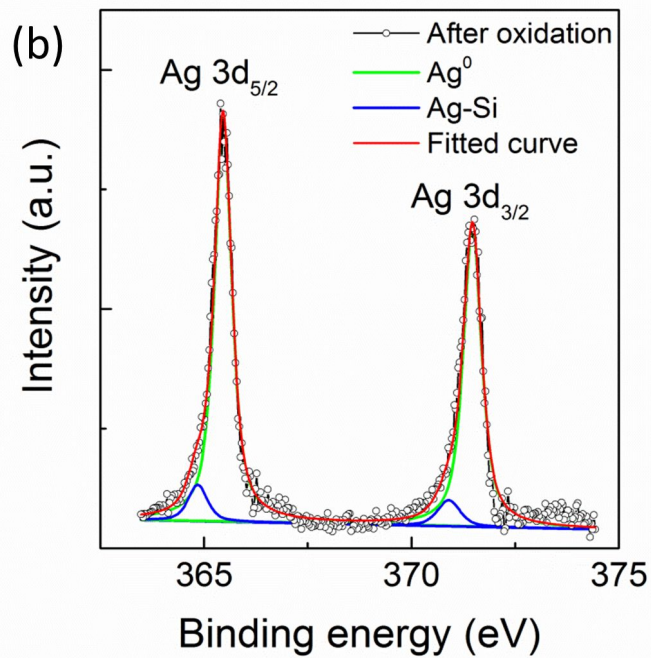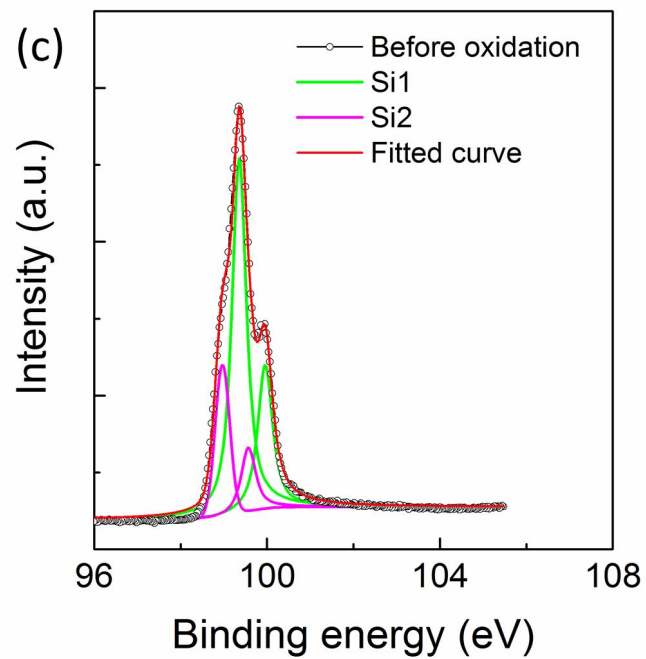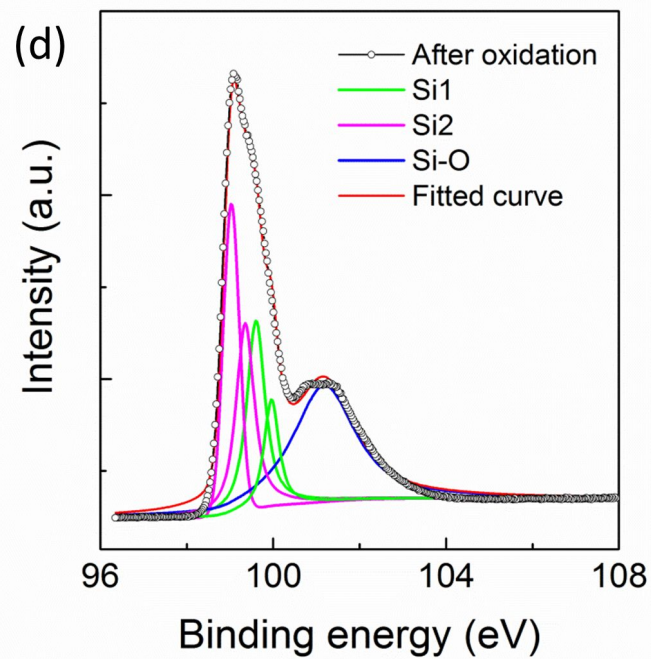

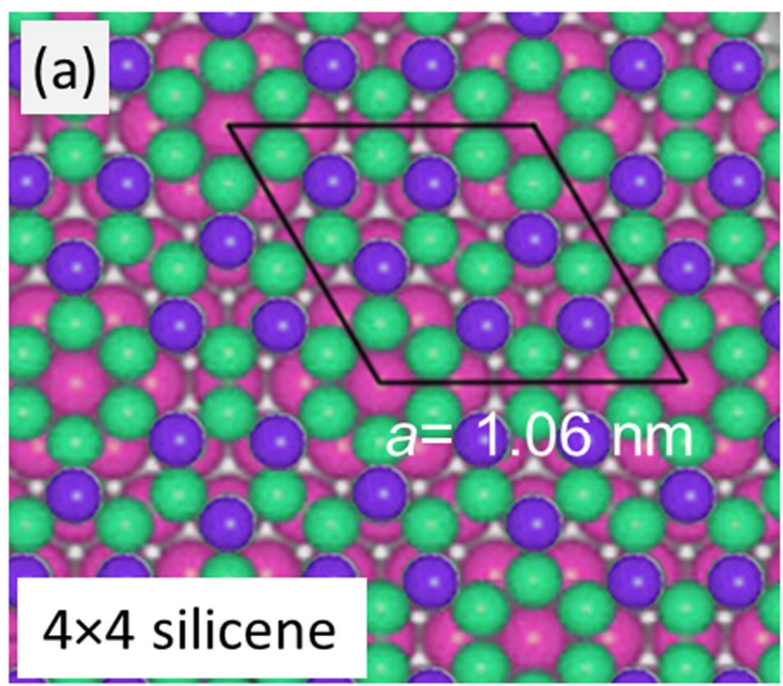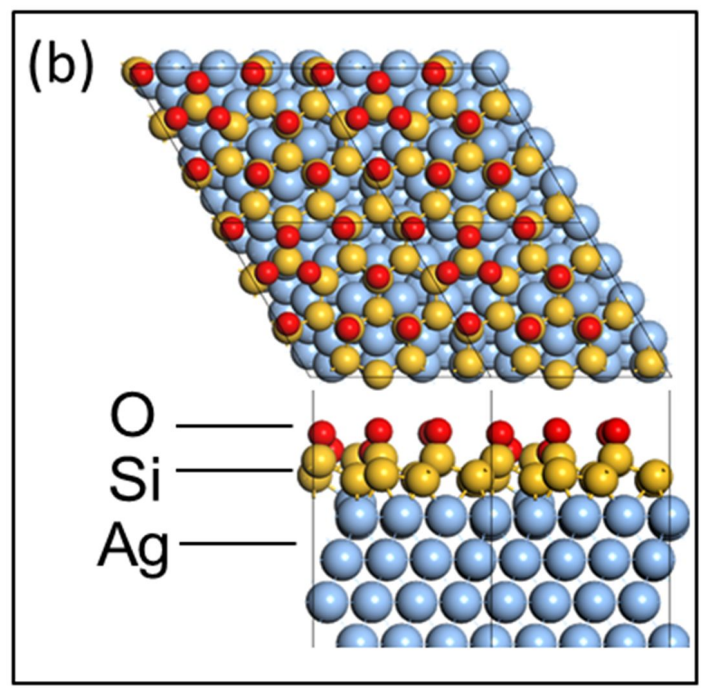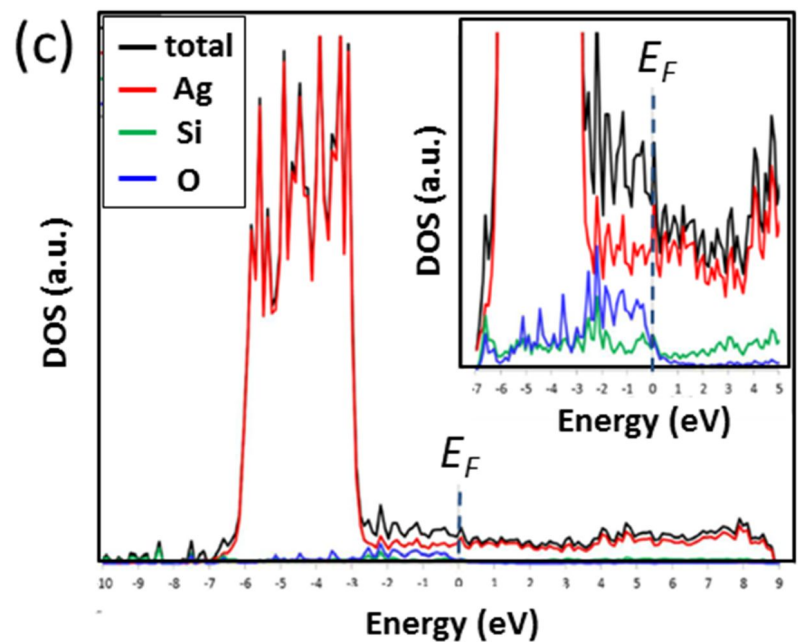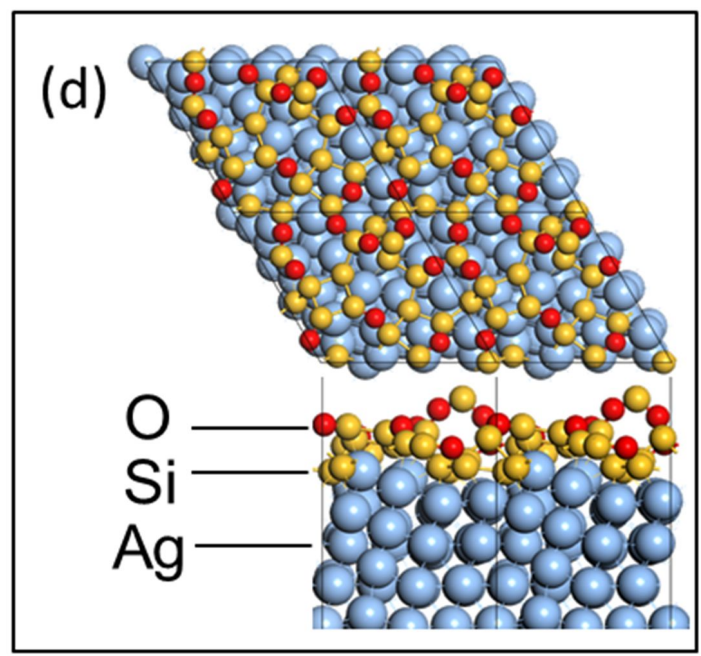